\documentclass[review]{elsarticle}
\usepackage{graphicx}
\usepackage{lineno,hyperref}
\modulolinenumbers[5]
\usepackage{epsfig}
\usepackage{longtable}
\journal{Phys. Lett. B}

\begin{document}

\begin{frontmatter}

\title{Production of multistrange hadrons, light nuclei and hypertriton in central Au+Au collisions at $\sqrt{s_{NN}}=$ 11.5 and 200 GeV}

\author{
N.~Shah$^{1}$,
Y.~G.~Ma$^{1}$,
J.~H.~Chen$^{1}$,
and~S.~Zhang$^{1}$
}
\address{$^{1}$Shanghai Institute of Applied Physics, Chinese Academy of Sciences, Shanghai 201800, China}


\begin{abstract}
The production of dibaryons, light nuclei and hypertriton in the most central Au+Au collisions at $\sqrt{s_{NN}}=$ 11.5 and 200 GeV are investigated by using a naive coalescence model.  The production of light nuclei is studied and found that the production rate reduces by a factor of 330 (1200) for each extra nucleon added to nuclei at  $\sqrt{s_{NN}}=$ 11.5 (200) GeV. The $p_{T}$ integrated yield of multistrange hadrons falls exponentially as strangeness quantum number increases. We further investigate strangeness population factor $S_{3}, S_{2}$ as a function of transverse momentum as well as $\sqrt{s_{NN}}$. The calculations for $\sqrt{s_{NN}}=$ 11.5 GeV presented here will stimulate interest to carry out these measurements during the phase-II of beam energy scan program at STAR experiment.
\end{abstract}


\end{frontmatter}
\date{November 16, 2015}%

\section{Introduction}

The experiments at Relativistic Heavy Ion Collider (RHIC) have shown evidence for the hot and dense matter, also known as quark gluon plasma (QGP), created during the early stages of the collisions~\cite{qgp1,qgp2,qgp3,qgp4}. The high temperature and baryon density of the produced matter makes it most suitable environment for the production of light nuclei (p, d, $^{3}He$, $^{4}He$), hypertriton and dibaryons ($\Lambda\Lambda$, p$\Omega$, $\Xi\Xi$, $\Omega\Omega$) as well as their antiparticles. 

For long time the study of light (anti)nuclei and (anti)hypernuclei production has remained of interest for physicists~\cite{ygm1,ygm2}. These studies are important to understand the matter-antimatter symmetry, dark matter and structure of neutron star~\cite{hori,ko}. Antihypertriton ($^{3}_{\bar\Lambda}\bar{H}$) and antihelium-4 ($^{4}\bar{He}$) have already been observed at RHIC~\cite{starscience, starnature}  and Large Hadron Collider~\cite{ALICE}. Very recently, interaction between antiproton pairs has been also measured by the STAR experiment~\cite{pbarpbar-star}.  The production of light (anti)nuclei and (anti)hypernuclei in heavy ion collisions is fairly described by the thermal model~\cite{therm1,therm2} and the coalescence model based on multiphase transport model as well as other transport models~\cite{zhang, chen, botvina, zhu}. The production of light (anti)nuclei and (anti)hypernuclei in the most central Au+Au collisions at $\sqrt{s_{NN}}=$ 200 GeV has been studied using coalescence model and hydrodynamic blast-wave model in~\cite{Xue, sun}. Using the same model of Ref.~\cite{Xue}, the production of light (anti)nuclei and (anti)hypernuclei in the most central Au+Au collisions at $\sqrt{s_{NN}}=$ 11.5 GeV are discussed in this article. 

Different quantum chromodynamics (QCD) based models have proposed existence of dibaryons as exotic form of matter. The H dibaryon was first predicted by Jaffe~\cite{jaffe} and then later many other dibaryon states were predicted, e.g.  p$\Omega$~\cite{pomega}, $\Xi\Xi$~\cite{xixi} and $\Omega\Omega$~\cite{omegaomega}. Recently experiments at RHIC~\cite{HSTAR} and LHC~\cite{HALICE} have searched for H dibaryon. With the advancement in computation power reasonable theoretical progress has been made to understand dibaryon structure~\cite{ExHIC1,ExHIC2,th-dib1,th-dib2,th-dib3}. However information about the invariant yield of dibaryons from heavy ion collisions remains scarce and more efforts are required in this direction. The invariant yield of dibaryons $\Lambda\Lambda$, p$\Omega$, $\Xi\Xi$ and $\Omega\Omega$ are presented for central Au+Au collisions at $\sqrt{s_{NN}}=$ 11.5 and 200 GeV.

The baryon-strangeness correlation coefficient $C_{BS}$ is proposed as a diagnostic tool to understand the nature of matter formed in heavy ion collisions~\cite{vkoch,haussler}. For QGP state the $C_{BS}$ is expected to be unity, however a significant dependence of $C_{BS}$ on hadronic environment is observed by V. Koch, A. Majumder and J. Randrup~\cite{vkoch}. Measurement of $C_{BS}$ in experiments is a technical challenge as one needs to measure baryon number and strangeness on event-by-event basis. Therefore the strangeness population factor $S_{3}$ was introduced by T. A. Armstrong {\it et al.}~\cite{armstrong}, which fairly depicts the local correlation between baryon number and strangeness~\cite{zhang}. Further we introduce $S_{2}$, which represents the local strangeness-strangeness correlations. Keeping in mind the technical challenges to measure $C_{BS}$, in this Letter, we concentrate on the strangeness population factor $S_{3}$, $S_{2}$ for central Au+Au collisions at $\sqrt{s_{NN}}=$ 11.5 and 200 GeV.   

\begin{figure*}
\begin{center}
\epsfxsize = 4.8in
\epsfysize = 3.5in
\epsffile{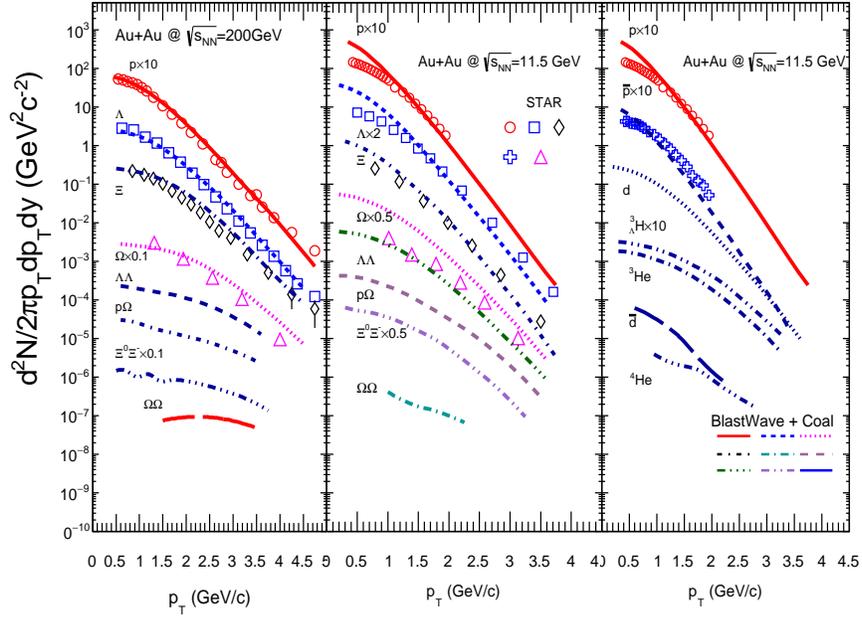} 
\end{center}
\caption{ (color online). Differential invariant yields versus p$_{T}$ distribution for p, $\Lambda$, $\Xi$, $\Omega$, $\Lambda\Lambda$, p$\Omega$, $\Xi^{0}\Xi^{-}$, $\Omega\Omega$, light (anti)nuclei and hypertriton produced in central Au+Au collisions at $\sqrt{s_{NN}} = 11.5$ and 200 GeV. The filled symbols are data from STAR experiment~\cite{star200_1,star200_2,star200_3,stardndy_11} and different lines represent our calculations from the hydrodynamical blast-wave model plus a coalescence model.}
\label{fig:spectra_11GeV}
\end{figure*}

\section{Coalescence Model}

A naive coalescence model is used to study the production of multistrange hadrons, light nuclei and hypertriton in central Au+Au collisions at $\sqrt{s_{NN}}=$ 11.5 and 200 GeV. It is assumed that the production of these particles occur at the kinetic freeze-out stage. In this case the particle production probability is proportional to the primordial hadron density and can be described by following equation~\cite{sato}:  
\begin{equation}
\label{EA}
E_{c} \frac{d^3N_{c}}{d^3p_{c}} =  B ( E_a \frac{d^3N_a}{d^3p_{a}})^{n} ( E_b \frac{d^3N_b}{d^3p_{b}})^{m},
\end{equation}
\noindent 
where E$\frac{d^3N}{d^3p}$ are the invariant yield of particles (a, b and c) under consideration, $p_{c}, p_{a}$ and $p_{b}$ are their momenta, B is the coalescence parameter and $\vec{p_{c}}=n\vec{p_{a}}+m\vec{p_{b}}$. The phase space information from the hydrodynamic blast-wave model is used as an input to the equation~(\ref{EA}) to calculate the invariant yields of $\Lambda\Lambda$, p$\Omega$, $\Xi^{0}\Xi^{-}$, $\Omega\Omega$, light nuclei and hypertriton. 

In hydrodynamic blast-wave model~\cite{bwm1}, the system is characterized by these parameters: the kinetic freeze-out temperature $T_{kin}$ , the radial flow parameter $\rho_0$ and elliptic flow parameter $\rho_2$ , the spatial anisotropy a, the average transverse radius R, and the particle emission duration $\tau_0$. It is assumed that the fireball created in heavy ion collision is in local thermal equilibrium and moves outward with velocity $u_{\mu}$. The phase-space emission points for hadrons are defined as a Wigner function:
\begin{eqnarray}
\label{sxp}
\nonumber
S(x,p) d^{4}x &= & \frac{2s+1}{(2\pi)^3}m_{t}\cosh(y-\eta)exp(-\frac{p^{\mu}u_{\mu}}{T_{k}}) \\
\nonumber
&& \times \Theta(1-\tilde{r}(r,\phi))H(\eta)\\
&& \times \delta(\tau-\tau_{0})d\tau\tau d\eta r drd\phi, 
\end{eqnarray}
\noindent 
where y is the rapidity,  $m_{t}$ is transverse mass, $p^{\mu}$ is four momentum, and (2s+1) is the degeneracy due to spin of hadrons. $\tilde{r}$ is given by

\begin{equation}
\label{rtilde}
\tilde{r} =\sqrt{\frac{(x^1)^2}{{R_{x}}^2}+\frac{(x^2)^2}{{R_{y}}^2}}, R_{x}=aR, R_{y}=\frac{R}{a},
\end{equation}
where $(x^{1},x^{2})$ is the transverse position of the hadrons in coordinate space. Then we can define the azimuthally integrated $p_{T}$ spectrum as

\begin{equation}
\label{dnptdpt}
\frac{dN}{2\pi p_{T}dp_{T}}=\int S(x,p)d^4x.
\end{equation}

Results obtained for the invariant yields of multistrange hadrons, nuclei and hypertriton using equation~\ref{EA} and~\ref{dnptdpt} are discussed in next section. 
\begin{table}
\scalebox{0.7}{
\begin{tabular}{ |c|| c| c| c| c| c| c| c|}
\hline
$\sqrt{s_{NN}}$  & $dN_{^{3}He}/dy$ & $dN_{^{3}_{\Lambda}H}/dy$ & $dN_{^{4}He}/dy$ &
$dN_{\Lambda\Lambda}/dy$ & $dN_{p\Omega}/dy$ & $dN_{\Xi^{0}\Xi^{-}}/dy$ & $dN_{\Omega\Omega}/dy$\\
(GeV) & & & & & & &\\
\hline
11.5 & $1.06\times10^{-2}$ & $2.04\times10^{-3}$   & $3.63\times10^{-5}$    &  $2.46\times10^{-2}$     & $2.12\times10^{-3}$  & $6.68\times10^{-4}$ &$1.63\times10^{-6}$ \\
\hline
200 & $1.65\times10^{-4}$ & $1.05\times10^{-4}$   & $3.30\times10^{-7}$    &  $7.24\times10^{-3}$     & $4.24\times10^{-4}$  & $2.75\times10^{-4}$ &$3.25\times10^{-6}$ \\
\hline
\end{tabular}}
\caption{\label{tab:int_dndy} $p_{T}$ integrated yields of light nuclei, hypertriton and dibaryons in Au+Au collisions. }
\end{table} 
\section{Result and discussion}

 To study dibaryons, light nuclei and hypertriton production in the central Au+Au collisions at the RHIC energies 200 GeV and 11.5 GeV, we use following parameters derived from the STAR experiment at the RHIC as input to the hydrodynamic blast-wave model: kinetic freeze-out temperature = 89 (120) MeV, baryo-chemical potential = 21.9 (315) MeV, strangeness chemical potential = 6.5 (68) MeV and radial flow parameter $\rho_0$ = 0.91 (0.46) for the $\sqrt{s_{NN}}=200$ (11.5) GeV~\cite{kumar}. The elliptic flow parameter $\rho_2$  = 0, spatial anisotropy a = 1, average transverse radius R = 10 fm and finite longitudinal proper time = 6.2 fm/c are set same for both $\sqrt{s_{NN}}=200 GeV$ and 11.5 GeV~\cite{rtau1,rtau2,rtau3}. Similar calculations were done by K.-J. Sun and L.-W. Chen in~\cite{sun}, where the freeze-out parameters are higher than the parameters used in our calculations. The proton spectra used from the PHENIX collaboration to derive the freeze-out parameter in~\cite{sun} are not corrected for the feed-down from $\Lambda$ and $\Sigma$ baryons.The coalescence of hadron occurs when $|\vec{r}_{i}-\vec{r}_{j}| < 2R_{0}$ and $|\vec{p}_{i}-\vec{p}_{j}| < 100$ MeV/c, where $(\vec{r}_{i},\vec{p}_{i})$ and $(\vec{r}_{j},\vec{p}_{j})$ are the phase-space position of the two constituent hadrons, and $R_{0}$ is the nuclear force radius. For deuteron and multistrange dibaryon $R_{0}$ = 1.57 fm is used and for the other nuclei $R_{0}$ = 1.5 fm is used.

The first two panels in figure~\ref{fig:spectra_11GeV} show differential yields of p, $\Lambda$, $\Xi$, $\Omega$, $\Lambda\Lambda$, p$\Omega$, $\Xi^{0}\Xi^{-}$ and $\Omega\Omega$ produced in central Au+Au collisions at $\sqrt{s_{NN}} = 200$ and 11.5 GeV respectively. Our calculation can reproduce the data for proton, $\Lambda$, $\Xi$ and $\Omega$ from the STAR experiment at both energies~\cite{star200_1,star200_2,star200_3,stardndy_11}. The light (anti)nuclei and hypertriton spectra for $\sqrt{s_{NN}} = 11.5$ GeV are shown in third panel of figure~\ref{fig:spectra_11GeV}. For $\sqrt{s_{NN}} = 200$ GeV the light (anti)nuclei spectra are taken from the article~\cite{Xue}, where same coalescence model was used. The $p_{T}$ integrated yield for light nuclei and dibaryons in the central rapidity are given in Table~\ref{tab:int_dndy}. We observe that expected yields of all the particles at $\sqrt{s_{NN}} = 11.5$ GeV are significantly higher than $\sqrt{s_{NN}} = 200$ except for $\Omega\Omega$, may be because of competition between strangeness production mechanism at this energy.

\begin{figure}
\begin{center}

\epsfxsize = 3.3in
\epsfysize = 2.5in
\epsffile{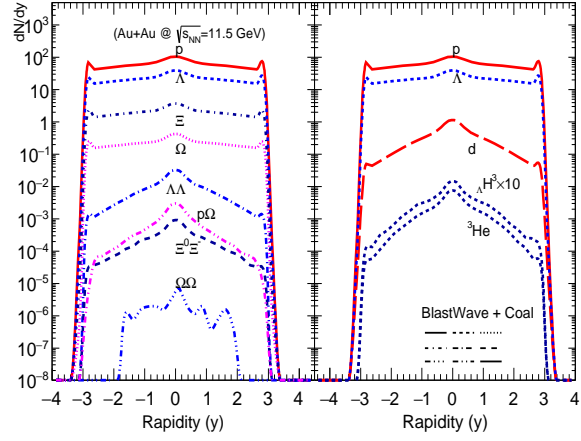} 
\end{center}
\caption{
(color online)  The rapidity distribution of multistrange hadrons, light nuclei and hypertriton in central Au+Au collisions at $\sqrt{s_{NN}} = 11.5$ GeV.}
\label{fig:rapidity}
\end{figure}

Figure~\ref{fig:rapidity} shows the rapidity distribution of p, $\Lambda$, $\Xi$, $\Omega$, $\Lambda\Lambda$, p$\Omega$, $\Xi^{0}\Xi^{-}$, $\Omega\Omega$, d, $^{3}He$ and hypertriton in central Au+Au collision at $\sqrt{s_{NN}} = 11.5$ GeV from the hydrodynamical blast-wave model plus a coalescence model. Since uniform rapidity distribution is used for $\sqrt{s_{NN}} = 200$ GeV, we have not shown those rapidity distributions here. 

\begin{figure}
\begin{center}
\epsfxsize = 3.5in
\epsfysize = 2.50in
\epsffile{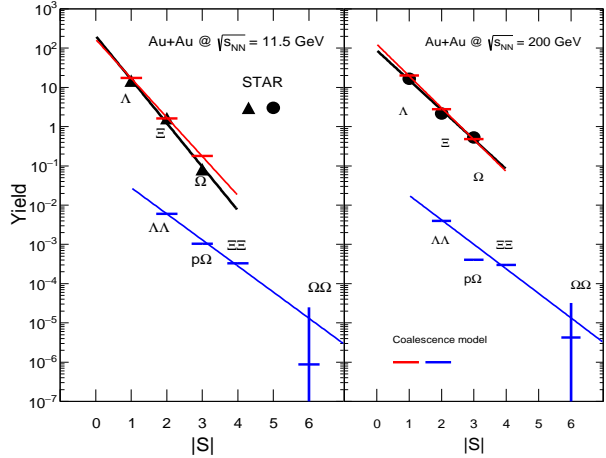} 
\end{center}
\caption{(color online) p$_{T}$ integrated yields {\it dN/dy} of multistrange hadrons as a function of strangeness $|S|$ for central Au+Au collisions at $\sqrt{s_{NN}} = 11.5$ GeV (left) and 200 GeV (right). The filled  symbols are data from the STAR experiment~\cite{star200_2,star200_3,stardndy_11}, the solid lines represent our calculations from the hydrodynamical blast-wave model plus a coalescence model and the dashed lines for the $\Lambda\Lambda$ and $\Omega\Omega$ dibaryons are from Ref~\cite{ExHIC2}.}
\label{fig:yieldvsS}
\end{figure}

The p$_{T}$ integrated yields {\it dN/dy} of multistrange hadrons as a function of strangeness $|S|$ for central Au+Au collisions at $\sqrt{s_{NN}} = 11.5$ GeV (left) and 200 GeV (right) are shown in figure~\ref{fig:yieldvsS}, where filled  symbols are data from the STAR experiment~\cite{star200_2,star200_3,stardndy_11} and different lines represent our calculations from the hydrodynamical blast-wave model plus a coalescence model.  The $\Lambda\Lambda$ and $\Omega\Omega$ dibaryon production yields at top RHIC energy were estimated by the ExHIC collaboration based on a realistic coalescence model and statistical model~\cite{ExHIC1,ExHIC2}. Those yields are compared with our calculations in the figure~\ref{fig:yieldvsS}. We observe an exponential behavior of the invariant yield of multistrange hadrons similar to light nuclei~\cite{Xue}. The yield for baryon and dibaryon systems are fitted with function $N_{S} = N^{i}({\frac{1}{\lambda}})^{|S|-1}$ , where $N^{i}$ is number of initial strange hadrons, $\lambda$ is penalty factor and $S$ is the strangeness. The penalty factor quantitatively tells us how hard it is to produce a hadron with strangeness ($|S|$+1) compared to a hadron with strangeness ($|S|$).  We obtain $\lambda$ = 9.86 for baryons and  $\lambda$ = 4.62 for dibaryon system from the model for central Au+Au collisions at $\sqrt{s_{NN}}=11.5$ GeV. Similarly we obtain $\lambda$ = 6.46 for baryons and  $\lambda$ = 4.21 for dibaryon system from the model for central Au+Au collisions at $\sqrt{s_{NN}}=200$ GeV. By fitting the data from the STAR experiment for baryons, we get  $\lambda$ = 12.92 $\pm$ 1.04 (5.71 $\pm$ 0.34) for the central Au+Au collisions at $\sqrt{s_{NN}}=11.5$ (200) GeV.

\begin{figure}
\begin{center}
\epsfxsize = 3.5in
\epsfysize = 2.50in
\epsffile{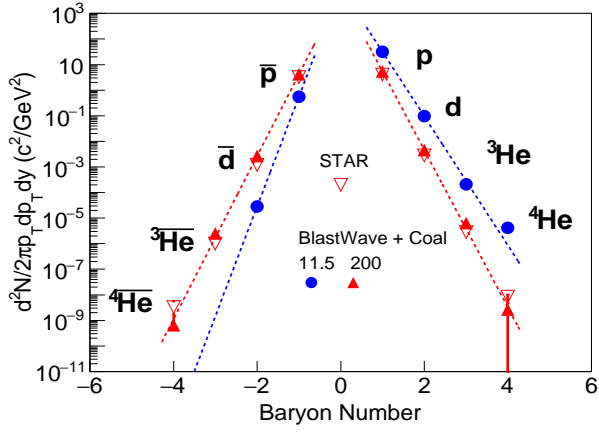} 
\end{center}
\caption{(color online) Invariant yield of nuclei in the average transverse momentum region ($p_{T}/|B|$ = 0.875 GeV/c) as a function of baryon number $B$ for central Au+Au collisions at $\sqrt{s_{NN}} = 11.5$ GeV and 200 GeV. Open symbols are data from STAR experiment~\cite{starnature}, solid points are our calculations from coalescence model and different lines represent fit to the coalescence model and data from STAR experiment~\cite{starnature}.}
\label{fig:yieldvsB}
\end{figure}

Figure~\ref{fig:yieldvsB} shows the production rate of nuclei as a function of baryon number for the central Au+Au collisions at $\sqrt{s_{NN}} = 11.5$ GeV and 200 GeV, where solid points are our calculations using the coalescence model and open  symbols are data from the STAR experiment~\cite{starnature}. At $\sqrt{s_{NN}} = 200$ GeV, our results are consistent with the STAR measurement within the uncertainties.  The production rates exhibit a decreasing exponential behavior with the increase in baryon number. Further we obtain the reduction factor by fitting the data with exponential function $e^{-rB}$.  Obtained fit values for reduction factor are 1.2$\times10^{3}$ (1.5$\times10^{3}$) and 0.33$\times10^{3}$ (1.95$\times10^{4}$)for adding one more nucleon (antinucleon) to the system for $\sqrt{s_{NN}} = 200$ and 11.5 GeV respectively. The reduction factor obtained from our calculation at  $\sqrt{s_{NN}} = 200$ GeV is comparable with $1.1^{+0.3}_{-0.2}\times10^{3}$ ($1.6^{+1.0}_{-0.6}\times10^{3}$) obtained by the STAR experiment~\cite{starnature}. The production rate for nuclei at  $\sqrt{s_{NN}} = 11.5$ GeV  are significantly higher than $\sqrt{s_{NN}} = 200$ GeV where the rates decrease sharply for antinuclei at same energy compared to $\sqrt{s_{NN}} = 200$ GeV. The difference in reduction factors between matter and antimatter shows a significant energy (or temperature) dependence, which illustrate an increasing matter-antimatter asymmetry of the yields at lower energies (temperatures). If we make a rough extension to current Universe at room temperature, we can hardly observe the antimatter existence, which is consistent with the current observation of the cosmic rays from which neither antideutron nor antihelium are observed~\cite{fuke, abe}. 

\begin{figure}
\begin{center}
\epsfxsize = 3.5in
\epsfysize = 2.50in
\epsffile{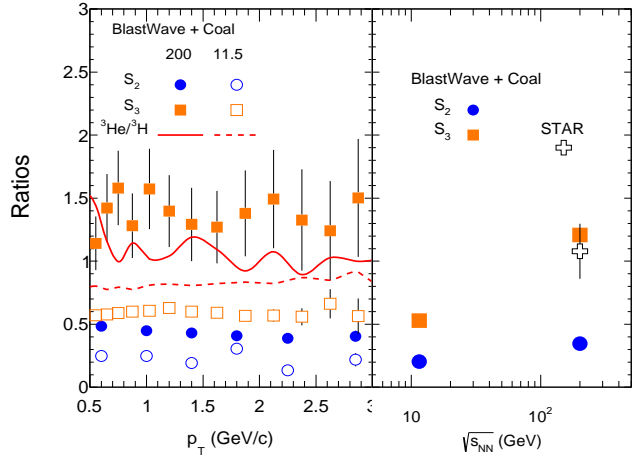} 
\end{center}
\caption{(color online)  On left ratios $S_{2}$, $S_{3}$ and $\frac{^{3}He}{^{3}H}$ are plotted as a function of transverse momentum ($p_{T}$) for central Au+Au collisions at $\sqrt{s_{NN}} = 11.5$ GeV and 200 GeV. On right ratios $S_{2}$ and $S_{3}$ are plotted as a function of beam energy $\sqrt{s_{NN}}$, where open cross is data from STAR experiment~\cite{starscience}.}
\label{fig:S3S2}
\end{figure}

The strangeness population factor $S_{3}=^{3}_{\Lambda}H/(^{3}He\times\frac{\Lambda}{p})$ contains the local baryon-strangeness correlation in the numerator and the baryon-baryon correlation in the denominator~\cite{zhang,sato}. Therefore $S_{3}$ is quantitatively a good representation of $\chi^{BS}_{11}/\chi^{B}_{2}$, where $\chi$ is the second derivative of the free energy with respect to the chemical potential,  from lattice QCD~\cite{cheng}. The ratio $S_{3}$ as a function transverse momentum is shown in figure~\ref{fig:S3S2} (left). Similarly we define $S_{2} = \Lambda\Lambda/(d\times(\Lambda/p)^{2})$ for strangeness=-2 dibaryon which contains the local strangeness-strangeness correlation in the numerator and the baryon-baryon correlation in the denominator.  An increase in ratios $S_{2}$ and $S_{3}$  is observed for Au+Au collisions at $\sqrt{s_{NN}}$ = 200 GeV compared to $\sqrt{s_{NN}}$ = 11.5 GeV as shown in figure~\ref{fig:S3S2} (right). The ratios $\frac{^{3}He}{^{3}H}$ for $\sqrt{s_{NN}} = 11.5$ GeV and 200 GeV are also shown in figure~\ref{fig:S3S2} (left). We observe that the ratio $\frac{^{3}He}{^{3}H}$  at $\sqrt{s_{NN}} = 11.5$ GeV is lower than unity, where the isospin effects become important compared to  $\sqrt{s_{NN}} =$ 200 GeV.  

\section{Conclusion}

We presented an interesting calculation for the production of dibaryons, light (anti)nuclei and hypertriton, based on a naive coalescence model for Au+Au collisions at $\sqrt{s_{NN}}$ = 11.5 and 200 GeV. The exponential behavior of the invariant yields versus strangeness is studied for the multistrange hadrons and penalty factor for the baryon and dibaryon are derived. The ratios $S_{2}$ and $S_{3}$ are discussed for Au+Au collisions at $\sqrt{s_{NN}}$ = 11.5 and 200 GeV. We observe an increase in $S_{2}$ and $S_{3}$ at $\sqrt{s_{NN}}$ = 200 GeV compared to $\sqrt{s_{NN}}$ = 11.5 GeV. 
Furthermore our study indicates that the suppression factor for nuclei production at $\sqrt{s_{NN}}$ = 11.5 GeV is roughly four times smaller than suppression factor at $\sqrt{s_{NN}}$ = 200 GeV; leading to higher probability for observation of light nuclei candidates at lower energy. Our calculation will provide the motivation to carry out measurement of $S_{3}$, light nuclei and dibaryons  during the phase-II of beam energy scan program at STAR experiment at RHIC~\cite{BES-2}.

\section{Acknowledgments}

This work is supported in part by the Major State Basic Research Development Program in China under Contract No. 2014CB845401, the National Natural Science Foundation of China under contract Nos. 11421505, 11520101004, 11275250, 11322547. Author N. Shah is  supported by Chinese Academy of Sciences (CAS) President's International Fellowship Initiative No. 2015PM029.

\section*{References}
\renewcommand{\bibfont}{\small}

%


\end{document}